
%

\newcount\driver \driver=1          

\newcount\mgnf\newcount\tipi\newcount\tipoformule
\mgnf=0          
\tipi=2          
\tipoformule=1   
\ifnum\mgnf=0
   \magnification=\magstep0\hoffset=0.cm
   \voffset=-0.5truecm\hsize=16.5truecm\vsize=24.truecm
   \parindent=4.pt\fi
\ifnum\mgnf=1
   \magnification=\magstep1\hoffset=0.truecm
   \voffset=-0.5truecm\hsize=16.5truecm\vsize=24.truecm
   \baselineskip=14pt plus0.1pt minus0.1pt \parindent=12pt
   \lineskip=4pt\lineskiplimit=0.1pt      \parskip=0.1pt plus1pt\fi
%
%
\let\a=\alpha   \let\g=\gamma    \let\d=\delta \let\e=\varepsilon
     \let\th=\vartheta \let\l=\lambda
    \let\n=\nu    \let\x=\xi           
\let\s=\sigma \let\t=\tau    
\let\ps=\psi  \let\o=\omega


{\count255=\time\divide\count255 by 60 \xdef\oramin{\number\count255}
        \multiply\count255 by-60\advance\count255 by\time
   \xdef\oramin{\oramin:\ifnum\count255<10 0\fi\the\count255}}

\def\ora{\oramin }

\def\data{\number\day/\ifcase\month\or gennaio \or febbraio \or marzo \or
aprile \or maggio \or giugno \or luglio \or agosto \or settembre
\or ottobre \or novembre \or dicembre \fi/\number\year;\ \ora}

\setbox200\hbox{$\scriptscriptstyle \data $}

\newcount\pgn \pgn=1
\def\foglio{\number\numsec:\number\pgn
\global\advance\pgn by 1}
\def\foglioa{A\number\numsec:\number\pgn
\global\advance\pgn by 1}

%


\global\newcount\numsec\global\newcount\numfor
\global\newcount\numfig
\gdef\profonditastruttura{\dp\strutbox}
\def\senondefinito#1{\expandafter\ifx\csname#1\endcsname\relax}
\def\SIA #1,#2,#3 {\senondefinito{#1#2}
\expandafter\xdef\csname #1#2\endcsname{#3} \else
\write16{???? ma #1,#2 e' gia' stato definito !!!!} \fi}
\def\etichetta(#1){(\veroparagrafo.\veraformula)
\SIA e,#1,(\veroparagrafo.\veraformula)
 \global\advance\numfor by 1
 \write16{ EQ \equ(#1) == #1  }}
\def \FU(#1)#2{\SIA fu,#1,#2 }
\def\etichettaa(#1){(A\veroparagrafo.\veraformula)
 \SIA e,#1,(A\veroparagrafo.\veraformula)
 \global\advance\numfor by 1
 \write16{ EQ \equ(#1) == #1  }}
\def\getichetta(#1){Fig. \verafigura
 \SIA e,#1,{\verafigura}
 \global\advance\numfig by 1
 \write16{ Fig. \equ(#1) ha simbolo  #1  }}
\newdimen\gwidth
\def\BOZZA{
\def\alato(##1){
 {\vtop to \profonditastruttura{\baselineskip
 \profonditastruttura\vss
 \rlap{\kern-\hsize\kern-1.2truecm{$\scriptstyle##1$}}}}}
\def\galato(##1){ \gwidth=\hsize \divide\gwidth by 2
 {\vtop to \profonditastruttura{\baselineskip
 \profonditastruttura\vss
 \rlap{\kern-\gwidth\kern-1.2truecm{$\scriptstyle##1$}}}}}
\footline={\rlap{\hbox{\copy200}\ $\st[\number\pageno]$}\hss\tenrm
\foglio\hss}
}
\def\alato(#1){}
\def\galato(#1){}
\def\veroparagrafo{\number\numsec}\def\veraformula{\number\numfor}
\def\verafigura{\number\numfig}
\def\geq(#1){\getichetta(#1)\galato(#1)}
\def\Eq(#1){\eqno{\etichetta(#1)\alato(#1)}}
\def\eq(#1){\etichetta(#1)\alato(#1)}
\def\Eqa(#1){\eqno{\etichettaa(#1)\alato(#1)}}
\def\eqa(#1){\etichettaa(#1)\alato(#1)}
\def\eqv(#1){\senondefinito{fu#1}$\clubsuit$#1\write16{No translation for #1}%
\else\csname fu#1\endcsname\fi}
\def\equ(#1){\senondefinito{e#1}\eqv(#1)\else\csname e#1\endcsname\fi}

\ifnum\tipoformule=1\let\Eq=\eqno\def\eq{}\let\Eqa=\eqno\def\eqa{}
\def\equ{{}}\fi

\def\include#1{
\openin13=#1.aux \ifeof13 \relax \else
\input #1.aux \closein13 \fi}
\openin14=\jobname.aux \ifeof14 \relax \else
\input \jobname.aux \closein14 \fi
%
%
%
\newdimen\xshift \newdimen\xwidth
%
%
\def\ins#1#2#3{\vbox to0pt{\kern-#2 \hbox{\kern#1 #3}\vss}\nointerlineskip}
%
%
\def\insertplot#1#2#3#4{
    \par \xwidth=#1 \xshift=\hsize \advance\xshift
     by-\xwidth \divide\xshift by 2 \vbox{
  \line{} \hbox{ \hskip\xshift  \vbox to #2{\vfil
 \ifnum\driver=0 #3  
                 \special{ps: plotfile #4.ps} 
 \fi
 \ifnum\driver=1  #3    \includegraphics{#4.ps}       \fi
 \ifnum\driver=2  #3   \ifnum\mgnf=0
                       \special{#4.ps 1. 1. scale}\fi
                       \ifnum\mgnf=1
                       \special{#4.ps 1.2 1.2 scale}\fi\fi
 \ifnum\driver=3 \ifnum\mgnf=0
                 \psfig{figure=#4.ps,height=#2,width=#1,scale=1.}
                 \kern-\baselineskip #3\fi
                 \ifnum\mgnf=1
                 \psfig{figure=#4.ps,height=#2,width=#1,scale=1.2}
                 \kern-\baselineskip #3\fi
 \ifnum\driver=5  #3  \fi
\fi}
\hfil}}}

\newskip\ttglue
\def\TIPI{
\font\ottorm=cmr8   \font\ottoi=cmmi8
\font\ottosy=cmsy8  \font\ottobf=cmbx8
\font\ottott=cmtt8  
\font\ottoit=cmti8
\def \ottopunti{\def\rm{\fam0\ottorm}
\textfont0=\ottorm  \textfont1=\ottoi
\textfont2=\ottosy  \textfont3=\ottoit
\textfont4=\ottott
\textfont\itfam=\ottoit  \def\it{\fam\itfam\ottoit}%
\textfont\ttfam=\ottott  \def\tt{\fam\ttfam\ottott}%
\textfont\bffam=\ottobf
\normalbaselineskip=9pt\normalbaselines\rm}
\let\nota=\ottopunti}
\def\TIPIO{
\font\setterm=amr7 
\font\settesy=amsy7 \font\settebf=ambx7 
\def \settepunti{\def\rm{\fam0\setterm}
\textfont0=\setterm   
\textfont2=\settesy   
\textfont\bffam=\settebf  \def\bf{\fam\bffam\settebf}
\normalbaselineskip=9pt\normalbaselines\rm
}\let\nota=\settepunti}

\def\TIPITOT{
\font\twelverm=cmr12
\font\twelvei=cmmi12
\font\twelvesy=cmsy10 scaled\magstep1
\font\twelveex=cmex10 scaled\magstep1
\font\twelveit=cmti12
\font\twelvett=cmtt12
\font\twelvebf=cmbx12
\font\twelvesl=cmsl12
\font\ninerm=cmr9
\font\ninesy=cmsy9
\font\eightrm=cmr8
\font\eighti=cmmi8
\font\eightsy=cmsy8
\font\eightbf=cmbx8
\font\eighttt=cmtt8
\font\eightsl=cmsl8
\font\eightit=cmti8
\font\sixrm=cmr6
\font\sixbf=cmbx6
\font\sixi=cmmi6
\font\sixsy=cmsy6
\font\twelvetruecmr=cmr10 scaled\magstep1
\font\twelvetruecmsy=cmsy10 scaled\magstep1
\font\tentruecmr=cmr10
\font\tentruecmsy=cmsy10
\font\eighttruecmr=cmr8
\font\eighttruecmsy=cmsy8
\font\seventruecmr=cmr7
\font\seventruecmsy=cmsy7
\font\sixtruecmr=cmr6
\font\sixtruecmsy=cmsy6
\font\fivetruecmr=cmr5
\font\fivetruecmsy=cmsy5
\textfont\truecmr=\tentruecmr
\scriptfont\truecmr=\seventruecmr
\scriptscriptfont\truecmr=\fivetruecmr
\textfont\truecmsy=\tentruecmsy
\scriptfont\truecmsy=\seventruecmsy
\scriptscriptfont\truecmr=\fivetruecmr
\scriptscriptfont\truecmsy=\fivetruecmsy
\def \eightpoint{\def\rm{\fam0\eightrm}
\textfont0=\eightrm \scriptfont0=\sixrm \scriptscriptfont0=\fiverm
\textfont1=\eighti \scriptfont1=\sixi   \scriptscriptfont1=\fivei
\textfont2=\eightsy \scriptfont2=\sixsy   \scriptscriptfont2=\fivesy
\textfont3=\tenex \scriptfont3=\tenex   \scriptscriptfont3=\tenex
\textfont\itfam=\eightit  \def\it{\fam\itfam\eightit}%
\textfont\slfam=\eightsl  \def\sl{\fam\slfam\eightsl}%
\textfont\ttfam=\eighttt  \def\tt{\fam\ttfam\eighttt}%
\textfont\bffam=\eightbf  \scriptfont\bffam=\sixbf
\scriptscriptfont\bffam=\fivebf  \def\bf{\fam\bffam\eightbf}%
\tt \ttglue=.5em plus.25em minus.15em
\setbox\strutbox=\hbox{\vrule height7pt depth2pt width0pt}%
\normalbaselineskip=9pt
\let\sc=\sixrm  \let\big=\eightbig  \normalbaselines\rm
\textfont\truecmr=\eighttruecmr
\scriptfont\truecmr=\sixtruecmr
\scriptscriptfont\truecmr=\fivetruecmr
\textfont\truecmsy=\eighttruecmsy
\scriptfont\truecmsy=\sixtruecmsy
}\let\nota=\eightpoint}

\newfam\msbfam   
\newfam\truecmr  
\newfam\truecmsy 
\newskip\ttglue
\ifnum\tipi=0\TIPIO \else\ifnum\tipi=1 \TIPI\else \TIPITOT\fi\fi

\def\didascalia#1{\vbox{\nota\0#1\hfill}\vskip0.3truecm}

%
\def\V#1{\vec#1}
\def\T#1{#1\kern-4pt\lower9pt\hbox{$\widetilde{}$}\kern4pt{}}

\let\dpr=\partial\let\io=\infty
\def\fra#1#2{{#1\over#2}}
\let\0=\noindent

\def\guida{\leaders\hbox to 1em{\hss.\hss}\hfill}
\def\tende#1{\vtop{\ialign{##\crcr\rightarrowfill\crcr
              \noalign{\kern-1pt\nointerlineskip}
              \hglue3.pt${\scriptstyle #1}$\hglue3.pt\crcr}}}
\def\otto{{\kern-1.truept\leftarrow\kern-5.truept\to\kern-1.truept}}

\def\acapo{\hfill\break}

\let\ciao=\bye

\def\etc{\hbox{\it etc}}\def\eg{\hbox{\it e.g.\ }}

\def\ie{\hbox{\it i.e.\ }}


\def\AA{{\V A}}\def\aa{{\V\a}}\def\nn{{\V\n}}\def\oo{{\V\o}}
\def\mm{{\V m}}\def\nn{{\V\n}}\def\lis#1{{\overline #1}}
\def\FF{{\cal F}}
\def\={{ \; \equiv \; }}

\def\Im{{\rm\,Im\,}}

\def\pps{{\V\ps{\,}}}

\def\2{{1\over2}}

\def\tst{\textstyle}
\def\st{\scriptscriptstyle}
\let\\=\noindent

\def\*{\vskip0.3truecm}

\catcode`\%=12\catcode`\}=12\catcode`\{=12
              \catcode`\<=1\catcode`\>=2

\openout13=gvnn.ps
\write13<
\write13<
\write13</punto { gsave 2 0 360 newpath arc fill stroke grestore} def>
\write13<0 90 punto     >
\write13<70 90 punto    >
\write13<120 60 punto   >
\write13<160 130 punto  >
\write13<200 110 punto  >
\write13<240 170 punto  >
\write13<240 130 punto  >
\write13<240 90 punto   >
\write13<240 0 punto    >
\write13<240 30 punto   >
\write13<210 70 punto   >
\write13<240 70 punto   >
\write13<240 50 punto   >
\write13<0 90 moveto 70 90 lineto>
\write13<70 90 moveto 120 60 lineto>
\write13<70 90 moveto 160 130 lineto>
\write13<160 130 moveto 200 110 lineto>
\write13<160 130 moveto 240 170 lineto>
\write13<200 110 moveto 240 130 lineto>
\write13<200 110 moveto 240 90 lineto>
\write13<120 60 moveto 240 0 lineto>
\write13<120 60 moveto 240 30 lineto>
\write13<120 60 moveto 210 70 lineto>
\write13<210 70 moveto 240 70 lineto>
\write13<210 70 moveto 240 50 lineto>
\write13<stroke>
\closeout13
\catcode`\%=14\catcode`\{=1
\catcode`\}=2\catcode`\<=12\catcode`\>=12


\vglue0.truecm


\0{\bf Non recursive proof of the KAM theorem}
@\footnote{${}^*$}{\nota Archived in
{\tt mp\_arc@math.utexas.edu}, \#93-229;
available also by e-mail request to the authors.}
\vskip.8truecm
\0{\bf Giovanni Gallavotti, Guido Gentile
\footnote{${}^1$}{\nota
Dipartimento di Fisica,
Universit\`a di Roma, ``La Sa\-pi\-en\-za", P.le Moro 2, 00185 Roma,
Italia. E-mail: {\tt gallavotti\%40221.} {\tt hepnet@lbl.gov},
{\tt gentileg\%39943.hepnet@lbl.gov}. } }
\vskip0.5truecm
\0{\bf Abstract:} {\sl A selfcontained proof of the KAM theorem in the
Thirring model is discussed, completely relaxing
the ``strong diophantine property'' hypothesis used in previous papers.}
\vskip0.5truecm
\0{\sl Keywords:\it\ KAM, invariant tori, classical mechanics,
perturbation theory, chaos}

\vskip1.3truecm

\\{\bf 1. Introduction}
\vglue.5truecm\numsec=1\numfor=1\pgn=1

\\In [G] a selfcontained proof of the KAM theorem in the Thirring model
is discussed, under the hypothesis that the rotation vectors $\oo_0$
verify a {\it strong diophantine property}.  At the end of the same
paper a heuristic argument is given to show that in fact such a
hypothesis can be relaxed.  In the present work we develop the
heuristic argument into an extension of the KAM theorem proof described
in [G]; the extension applies to rotations vectors verifying only the
usual diophantine condition.  It is a proof again based on Eliasson's
method, [E].

In our opinion this shows that a hypothesis like the strong diophantine
property of [G], or something similar to it, is very natural as it
simplifies the structure of the proof by separating from it the analysis
of a simple arithmetic property, whose untimely analysis would obscure
the proof.

For an introductory discussion of the model and a
more organic exposition of the problem, we refer to [G], [G1],
and to the references there reported. In the remaining
part of this section we confine ourselves to define the
model, to introduce the basic notation, and to give the
result we have obtained.

The Thirring model, [T], is described by the hamiltonian:
$$ \fra12 J^{-1}\AA\cdot\AA\,+\,\e f(\aa) \Eq(1.1) $$
where $J$ is the (diagonal) matrix of the inertia moments,
$\AA=(A_1,\ldots,A_l)\in R^l$ are their angular momenta and
$\aa=(\a_1,\ldots,\a_l)\in T^l$ are the angles describing their
positions: the matrix $J$ will be supposed non singular; but we only
suppose that $\min_{j=1,\ldots,l}J_j=J_0>0$, and no assumption is
made on the size of the {\it twist rate} $T=\min J_j^{-1}$: the results
will be uniform in $T$ (hence they can be called ``twistless
results''). We
suppose $f$ to be an even trigonometric polynomial of degree $N$:
$$ f(\aa)=\sum_{0<|\nn|\le N} f_\nn\,\cos\nn\cdot\aa, \qquad \qquad
f_\nn=f_{-\nn} \; , \qquad |\nn| = \sum_{j=1}^l |\nn_j| \Eq(1.2) $$
We shall consider a ``rotation vector'' $\oo_0=(\o_1,
\ldots,\o_l)\in R^l$ verifying the {\it diophantine condition}:
$$ \bar C_0|\oo_0\cdot\nn|\ge |\nn|^{-\t}, \kern1.5cm\V0\ne\nn\in
Z^l \Eq(1.3) $$
with diophantine constants $\bar C_0, \t$. The {\it diophantine
vectors} have full measure in $R^l$ if $\t$ is fixed $\t>l-1$.
We shall set $\AA_0=J\oo_0$.

As in [G], we prove the following result.

\*

\\{\bf Theorem}:{\it The system described by the Hamiltonian
\equ(1.1) admits an $\e$--analytic family of motions starting at
$\aa=\V0$ and having the form:
$$ \AA = \AA_0 + \V H(\oo_0t;\e),\qquad\aa=
\oo_0t+\V h(\oo_0 t;\e) \Eq(1.4) $$
with $\V H(\pps;\e),\V h(\pps;\e)$  analytic, divisible by $\e$,
for $|\Im \psi_j|<\x$, $\pps\in T^l$, and for $|\e|<\e_0$ with:
$$ \e_0^{-1} = b J_0^{-1} (2^\t \bar C_0 )^2 f_0 N^{2+l}
e^{c N}e^{\x N} \Eq(1.5) $$
where $b,c$ are $l$--dependent positive constants,
$f_0=\max_\nn |f_\nn|$.}

\*

This means that the set $\AA=\AA_0+\V H(\pps;\e), \, \aa=\pps+\V
h(\pps;\e)$ described as $\pps$ varies in $T^l$ is, for $\e$ small
enough, an invariant torus for \equ(1.1), which is run quasi
periodically with angular velocity vector $\oo_0$. It is a family of
invariant tori coinciding, for $\e=0$, with the torus
$\AA=\AA_0,\,\aa=\pps\in T^l$.
The presence of the factor $2^\t$
marks the only difference from the analogous result in [G].

Calling $\V H^{(k)}(\pps),\V h^{(k)}(\pps)$ the $k$-th order
coefficients of the Taylor expansion of $\V H,\V h$ in powers of $\e$
and writing the equation of motion as $\dot\aa=J^{-1}\AA$ and
$\dot\AA=-\e\dpr_\aa f(\aa)$ we get immediately recursion
relations for $\V H^{(k)},\V h^{(k)}$, namely, for $k>1$:
$$ \eqalign{
{\V\o}_0 \cdot\V\dpr\,h^{(k)}_j & =
J^{-1}_j H^{(k)}_j \cr
{\V\o}_0\cdot\V\dpr\,H^{(k)}_j & =
-  \sum_{m_1,\ldots,m_l\atop|\mm|>0}\fra1{\prod_{s=1}^l m_s!}
\dpr_{\a_j}\, \dpr^{m_1+\ldots+m_l}_{\a_1^{m_1}\ldots\a_l^{m_l}}
f(\oo_0 t) \cdot {\sum}^* \prod_{s=1}^l\prod_{j=1}^{m_s}
h^{(k^s_j)}_s(\oo_0 t) \cr} \Eq(1.6) $$
where the $\sum^*$ denotes summation over the integers $k^s_j\ge1$
with: $\sum_{s=1}^l\sum_{j=1}^{m_s}k^s_j=k-1$.

The trigonometric polynomial $\V h^{(k)}(\pps)$ will be completely
determined (if possible at all) by requiring it to have $\V0$ average
over $\pps$, (note that $\V H^{(k)}$ has to have zero average over
$\pps$).  For $k=1$ it is:
$$ \tst\V h^{(1)}(\pps)=-\sum_{\nn\ne\V0}
\fra{iJ^{-1}\nn}{(i\oo_0\cdot\nn)^2}
f_\nn\,e^{i\nn\cdot\pps} \Eq(1.7) $$

One easily finds that
the equation for $\V h^{(k)}$ can be solved
and its solution is a trigonometric polynomial
in $\pps$, of degree $\le k N$, odd if $\V
h^{(k)}$ is determined by imposing that its average over $\pps$
vanishes.

The remaining part of the paper is structured as follows:
in section 2 we set a diagrammatic expansion of
$\V h^{(k)}$, as in [G]. In section 3 we discuss
a proposition which leads to the original result of this
paper, and in section 4 we prove the theorem,
repeating the discussion in [G], with some minor
changes due to the weakening of the strong
diophantine property hypothesis.

The above theorem fully reproduces, in the model \equ(1.1),
the theorem of Eliasson: for another alternative proof
of the same theorem with no assumption of parity or of
finite degree on the trigonometric
polynomial $f$, see [CF].

\vskip1.truecm

\\{\bf 2. Diagrammatic expansion}
\vglue.5truecm\numsec=2\numfor=1\pgn=1

\\Let $\th$ be a tree diagram: it will consist of a family of
``lines'' (\ie segments) numbered from $1$ to $k$ arranged to form
a (rooted) tree diagram as in the figure:
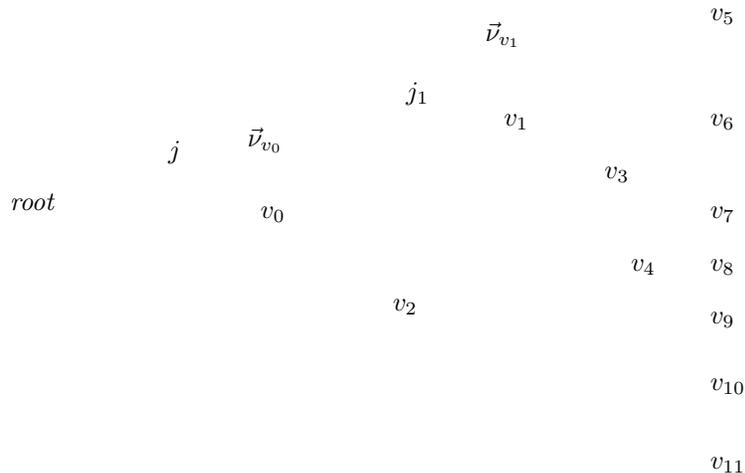
\insertplot{240pt}{170pt}{
\ins{-35pt}{90pt}{\it root}
\ins{25pt}{110pt}{$j$}
\ins{60pt}{85pt}{$v_0$}
\ins{55pt}{115pt}{$\nn_{v_0}$}
\ins{115pt}{132pt}{$j_{1}$}
\ins{152pt}{120pt}{$v_1$}
\ins{145pt}{155pt}{$\nn_{v_1}$}
\ins{110pt}{50pt}{$v_2$}
\ins{190pt}{100pt}{$v_3$}
\ins{230pt}{160pt}{$v_5$}
\ins{230pt}{120pt}{$v_6$}
\ins{230pt}{85pt}{$v_7$}
\ins{230pt}{-10pt}{$v_{11}$}
\ins{230pt}{20pt}{$v_{10}$}
\ins{200pt}{65pt}{$v_4$}
\ins{230pt}{65pt}{$v_8$}
\ins{230pt}{45pt}{$v_9$}
}{gvnn}
\kern1.3cm
\didascalia{fig. 1: A tree diagram $\th$ with
$m_{v_0}=2,m_{v_1}=2,m_{v_2}=3,m_{v_3}=2,m_{v_4}=2$ and $m=12$,
$\prod m_v!=2^4\cdot6$, and some decorations. The line numbers,
distinguishing the lines, are not shown.}

To each vertex $v$ we attach a ``mode label''
$\nn_v\in Z^l,\,|\nn_v|\le N$ and to each branch leading to $v$ we
attach a ``branch label'' $j_v=1,\ldots,l$.
The order of the diagram will be $k=$ number of vertices $=$
number of branches (the tree root will not be regarded as a vertex).

We imagine that all the diagram lines have the same length (even though
they are drawn with arbitrary length in fig.1).  A group acts on the
set of diagrams, generated by the permutations of the subdiagrams having
the same vertex as root.  Two diagrams that can be superposed by the
action of a transformation of the group will be regarded as identical
(recall however that the diagram lines are numbered, \ie are regarded as
distinct, and the superpositon has to be such that all the decorations
of the diagram match).  Tree diagrams are regarded as partially ordered
sets of vertices (or lines) with a minimal element given by the root (or
the root line). We shall imagine that each branch carries also an arrow
pointing to the root (``gravity'' direction, opposite to the order).

We define the ``momentum'' entering $v$ as $\nn(v)=\sum_{w\ge v}\nn_w$:
therefore the momentum entering a vertex $v$ is given
by the sum of the momenta entering the immediately following vertices
plus the ``momentum emitted'' by $v$ (\ie the mode $\nn_v$).
If from a vertex $v$ emerge $m_1$ lines carrying a label $j=1$, $m_2$
lines carrying $j=2$, $\ldots$, it follows that \equ(1.6) can be
rewritten:
$$ h^{(k)}_{\nn j}=\fra1{k!}
{\sum}^*\prod_{v\in\th}\fra{(-i J^{-1}\nn_v)_{j_v}\,
f_{\nn_v}\prod_{s=1}^l(i\nn_v)^{m_s}_s}{(i\oo_0\cdot\nn(v))^2}
\Eq(2.1) $$
with the sum running over the diagrams $\th$ of order $k$ and with
$\nn(v_0)=\nn$; and the combinatorics can be checked from \equ(1.6), by
taking into account that we regard the diagram lines as all different
(to fix the factorials).
The ${}^*$ recalls that the diagram $\th$ can and will be supposed such
that $ \nn(v) \ne \V0 $ for all $ v\in\th $ (by the remarked parity
properties of $\V h^{(k)}$).

As in [G], according to
Eliasson's terminology, we can define the {\it resonant diagrams}
as the diagrams with vertices $v',v$, with $v'<v$, not necessarily
nearest neighbours, such that $\nn(v)=\nn(v')$.
If there were no resonant diagrams, it would be straightforward
to obtain a bound like \equ(1.5), as it is shown in [G].
However {\it there are} resonant diagrams.The key remark is that they
cancel {\it almost exactly}. For a more detailed heuristic discussion
about the two above remarks we refer again to [G].

The tree diagrams will play the role of {\it Feynman
diagrams} in field theory; and they will be plagued by
overlapping divergences. They
will therefore be collected into another family of graphs,
that we shall call {\it trees}, on which the bounds are easy.
The $(\oo\cdot\nn)^{-2}$ are the {\it propagators}, in our analogy.

In [G], a {\it scaling} parameter $\g$ (which could be taken $\g=2$)
was fixed. Then, defining the adimensional frequency
$\oo\=\bar C_0\oo_0$, a propagator $(\oo\cdot\nn)^{-2}$ was said to be
{\it on scale $n$} if $2^{n-1}<|\oo\cdot\nn|\le
2^n$, for $n\le0$, and it was set $n=1$ if $1<|\oo\cdot\nn|$.
Nevertheless, if we want to eliminate the strong diophantine
property, we need to change the decomposition of the propagator.
We define a new vector $\oo= 2^\t \bar C_0 \oo_0$, and we say
that a propagator $(\oo\cdot\nn)^{-2}$ is
{\it on scale $n$} if $\g_{n-1}<|\oo\cdot\nn|\le
\g_n$, for $n\le0$, and we set $n=1$ if $\g_0<|\oo\cdot\nn|$.
Here $\{ \g_n \}$ is a sequence such that $1/2 \le \g_n 2^{-n}
\le 1 $, which will be suitably chosen:
how to fix such a sequence will be explained in section 3.
Here we outline that the fixing of the sequence depends
on the rotation vector we have chosen:
the conceptual advantage we have,
with respect to the result obtained under
the strong diophantine property hypothesis, is that the sequence of
scales $\g_n$ is not ``prescribed a priori'' but it is
determined by the arithmetic properties of the rotation
vector $\oo_0$ (see the ending comments in [G]).

Proceeding as in quantum field theory, see [G3], given a diagram $\th$,
we can attach a {\it scale label} to each line $v'v$ in \equ(8) (with
$v'$ being the vertex preceding $v$): it is
equal to $n$ if $n$ is the scale of the line propagator.
Note that the labels thus attached to a diagram are uniquely
determined by the diagram: they will have only the function of
helping to visualize the orders of magnitude of the various diagram
lines.

Looking at such labels we identify the connected clusters $T$ of
vertices that are linked by a continuous path of lines with the same
scale label $n_T$ or a higher one.  We shall say that {\it the cluster
$T$ has scale $n_T$}.
As far as we are concerned, we can visualize a tree as a
collection of clusters between which there exists a relation
of partial ordering.

Among the clusters we consider the ones with the property that there is
only one diagram line entering them and only one exiting and both carry
the same momentum. Here we use that the diagram lines carry an arrow
pointing to the root: this gives a meaning to the words ``incoming'' and
``outgoing''.

If $V$ is one such cluster we denote $\l_V$ the incoming line: the line
scale $n=n_{\l_V}$ is smaller than the smallest scale $n'=n_V$ of the
lines inside $V$.  We call $w_1$ the vertex into which the line $\l_V$
ends, inside $V$.  {\it We say that such a $V$ is a {\it resonance} if
the number of lines contained in $V$ is $\le E\,2^{-n\e}$}, where
$n=n_{\l_V}$, and $E,\e$ are defined by:
$E\=2^{-3\e}N^{-1},\,\e=\t^{-1}$.  We call $n_{\l_V}$ {\it
the resonance scale}, and $\l_V$ a {\it resonant line}.

\vskip1.truecm

\\{\bf 3. Multiscale decomposition of the propagator}
\vglue.5truecm\numsec=2\numfor=1\pgn=1

\\Given $\oo_0$ verifying \equ(1.3) with some constant $\bar C_0$ and some
$\t>0$ we define $C_0\=2^\t \bar C_0$: this leaves \equ(1.3) still valid.
We define the set $B_n$ of the values of $|\oo\cdot\nn|$ as $\nn$ varies
in the set $0\le|\nn|< (2^{n+3})^{-1/\t}$, with $n=0,-1,-2,\ldots$. The
sets $B_n$ verify the inclusion relation $B_n\subset B_m$ if $m<n$.
The main property of the sets $B_n$ is that the spacing between their
elements is at least $2^\t(2(2^{n+3})^{-1/\t})^{-\t}\ge 2^{n+3}$, by
the diophantine property \equ(1.3); also $x\in B_n$ is such that
$x>2^{n+3}$, if $x\ne0$.

More abstractly let $B_n$, $n=0,-1,\ldots$, be a sequence of sets such
that i) $0\in B_n$, ii) $B_n\subset B_m$ if $m<n$ and iii) the spacing
between the points in $B_n$ is at least $2^{n+3}$ (the latter will be
the {\it spacing property}).  Then we can prove the following lemma:
\*
\0{\bf Lemma:} {\it There exists a sequence $\g_0,\g_{-1},\ldots$ with
$\g_p\in[2^{p-1},2^p]$ such that:
$$|x-\g_p|\ge 2^{n+1},\kern2.cm {\rm if}\quad n\le p\le0
\quad\quad {\rm and}\quad x \in B_n
\Eq(3.1)$$
for all $n \le 0$.}
\*
\\{\it Remark:} hence if $x\in \cup_n B_n$ with $|x|\le \g_p$ then
$|x|<\g_p$.
\*
\0{\it proof:} Note that if $\g_p\in [2^{p-1},2^p]\=I_p$ the \equ(3.1) are
obviously verified for $n>p-3$, hence we can suppose $p\ge n+3$.

\def\PP{{\cal P}}
Fix $p\le0$ and let $G=[a,b]$ be an interval verifying what we shall
call below the {\it property $\PP_n$}:
$$\PP_n\,:\kern1.cm
|x-\g|\ge 2^{m+1}\quad {\rm for\ all}\quad n\le m\le-3,\qquad \g\in G,
\qquad
|G|\ge 2^{n+1}\Eq(3.2)$$
Let $G_{p-3}\=[2^{p-1},b_{p-3}]$, with $b_{p-3}\in [2^{p-1},2^p]$, be a
{\it maximal} interval verifying property $\PP_{p-3}$
(note that $G_{p-3}$ exists because $x\in B_{p-3},\,x\ne0$
implies $x\ge 2^p$ by the spacing property, and it
is $b_{p-3} \ge 2^{p-1} + 2^{p-2}$).

Assume inductively that the intervals $[a_n,b_n]=G_n$ can be so chosen that
$G_{n'}\subseteq G_{n\prime\prime}$ if $n'<n\prime\prime$ and
$G_n$ is maximal among the intervals contained in $G_{n+1}$ and
verifying the property $\PP_n$.

If we can check that the hypothesis implies the existence of an interval
$G\subseteq G_n$ verifying $\PP_{n-1}$ we shall be able to define
$G_{n-1}$ to be a maximal interval among the ones contained in $G_n$
and with the property $\PP_{n-1}$: in case of ambiguity we shall take
$G_{n-1}$ to coincide with the rigthmost possible choice.

Clearly we shall be able to define $\g_p=\lim_{n\to-\io} b_n$, which
will verify \equ(3.1).

To check the existence of $G\subseteq G_n$ verifying $\PP_{n-1}$ we
consider first the case in which $B_{n-1}$ has one and only one point
$x$ in $G_n$.  If $|G_n|\ge 2^{n+2}$ and $x$ is in the first half of
$G_n$ we can take, by the spacing property,
$G=[x+2^n,\min\{b_n,\,x+2^{n+2}-2^n\}]$; if it is in the second half
we take $G=[\max\{a_n,x-2^{n+2}+2^n\},x-2^n]$.

If, on the other hand, $|G_n|<2^{n+2}$ it is $G_n\subset G_{n+1}$
strictly and, furthermore, the interval $(x-2^{n+2},x+2^{n+2})$ does not
contain points of $B_{n-1}$ other than $x$ itself (by the spacing
property).  The strict inclusion implies that there is a point $y\in
B_n$ at distance exactly $2^{n+1}$ from $G_n$ (recall the maximality of
$G_n$).

Suppose that $x$ is in the first half of $G_n$ and $y<a_n$, \ie
$y=a_n-2^{n+1}$; then $x-y< 2^{n+1}+2^{n+1}$ contradicting the spacing
property.  Hence $y>b_n$, \ie $y=b_n+2^{n+1}$: in such case it cannot
be, again by the spacing property, that $x+2^{n+2}>y=b_n+2^{n+1}$, so
that $b_n-x\ge 2^{n+1}$ and we can take $G=[b_n-2^n,b_n]$.  If $x$ is
in the second half the roles of left and right are exchanged.

This completes the analysis of the case in which only one point of
$B_{n-1}$ falls in $G_n$. The cases in which either no point or at least two
points of $B_n$ fall in $G_n$ are analogous but easier.
\footnote{${}^2$}{\nota
If two consecutive points $x<y$ of $B_{n-1}$ fall inside $G_n$ we must
have $y-x\ge 2^{n+2}$ by the spacing property: hence
$G=[x+2^n,y-2^n]\subset G_n$ enjoys the property $\PP_{n-1}$.
If no point of $B_{n-1}$ falls in $G_n$ let $y\in B_{n-1}$ be the
closest point to $G_n$; if its distance to $G_n$ exceeds $2^n$ we take
$G=G_n$. Otherwise suppose that $y>b_n$: the spacing property implies
that the interval $(y-2^{n+2},y)$ is free of points of $B_{n-1}$. Hence
if $a=\max\{a_n,y-2^{n+2}+2^n\}$, $b=y-2^n$ then $G=[a,b]$ has the
property $\PP_{n-1}$. If, instead, $y<a_n$ we set $a=y+2^n>a_n$ and
$b=\min\{b_n,y+2^{n+2}-2^n\}$ and $G=[a,b]$ enjoys property $\PP_{n-1}$.
}
\*

\0{\it Remark}: note that the above proof is constructive.

\*
Consider the special case in which the sets $B_n$ are the ones defined at
the beginning of the section. The above lemma can then be translated
into the following arithmetic proposition.
\*
\0{\bf Proposition}: {\it Given a diophantine vector $\oo_0$, \ie a vector
verifying \equ(1.3), let $C_0=2^\t \bar C_0$ and $\oo= C_0\oo_0$; it is
possible to find a sequence $\g_p\in[2^{p-1},2^p]$ such that
$\g_{p-1}\le\g_p$ and:
$$\big||\oo\cdot\nn|-\g_p\big|\ge 2^{n+1},\qquad{\rm if}\quad 0<|\nn|\le
(2^{n+3})^{-\t^{-1}}\Eq(3.3)$$
for all $n\le0$ and for all $p\ge n$. Furthermore
$|\oo\cdot\nn|\ne\g_n$, for all $n\le0$.}
\*
{\it Remark:}

1) The sequence $\g_p$ is constructively defined by the proof
above.

2) The \equ(3.3) is very similar to the condition added in [G] to the
\equ(1.3) to define the {\it strong diophantine property}.  The point
of the present paper is that all that is really needed (see the
following \S4) to prove the theorem in \S1 are \equ(1.3) and \equ(3.3):
the latter is a simple arithmetic property which is in fact a
consequence of \equ(1.3).

3) As remarked in [G] almost all $\oo_0$ verify \equ(1.3) for some
$\bar C_0$ and some $\t>l-1$ with a sequence $\g_p$ that can be
prescribed {\it a priori} as $\g_p=2^p$.  This, however, leaves out
important cases like $l=2$ and $\oo_0$ with a quadratic irrational as
rotation number.  And it has the very unfortunate drawback of being non
constructive, as the set of full measure of the $\oo$ verifying the
strong diophantine property is constructed by abstract nonsense
arguments (\eg the Borel Cantelli lemma).  Nevertheless considering
strongly diophantine vectors is natural as it leads to a simplified
proof, [G], of the KAM theorem, with Eliasson's method, eliminating one
side difficulty.

4) For the purpose of comparison note that the final comment of ref [G]
conjectures the above proposition: however the constant $C_0$
introduced there is $2 C_0$ in the above notations and, therefore, the
quantities called there $\g_p$ are $2$ times the ones in \equ(3.3) (in
other words the first inequality in \equ(3.3) has to be multiplied
side by side by $2$ to become the statement of [G]).

\vskip1.truecm

\\{\bf 4. Proof of the theorem}
\vglue.5truecm\numsec=2\numfor=1\pgn=1

\\Let us consider a diagram $\th$ and its clusters.  We wish to estimate
the number $N_n$ of lines with scale $n\le0$ in it, assuming $N_n>0$
(we remind that a line is on scale $n$, if the line propagator
$(\oo \cdot \nn)^{-2}$ is such that
$\g_{n-1} < |\oo \cdot \nn| < \g_n$, see also the remark
after \equ(3.1)).

Denoting $T$ a cluster of scale $n$ let $m_T$ be the number of
resonances of scale $n$ contained in $T$ (\ie with incoming lines of
scale $n$); we have the following inequality, valid for any diagram
$\th$:
$$ N_n\le\fra{4k}{E\,2^{-\e n}}+\sum_{T, \,n_T=n}
(-1+m_T)\Eq(4.1) $$
with $E=N^{-1}2^{-3\e},\e=\t^{-1}$. This is an extension of Brjuno's
lemma, [B], [P], called in [G] ``resonant
Siegel-Brjuno bound'': the proof, extracted from [G],
can be found in appendix.

Consider a diagram $\th^1$; we define the family $\FF(\th^1)$ generated
by $\th^1$ as follows. Given a resonance $V$ of $\th^1$ we detach the
part of $\th^1$ above $\l_V$ and attach it successively to the points
$w\in\tilde V$, where $\tilde V$ is the set of vertices of $V$
(including the endpoint $w_1$ of $\l_V$ contained in $V$) outside the
resonances contained in $V$. We say that a line is in $\tilde V$, if
it is contained in $V$ and has at least one point in $\tilde V$.
Note that all the lines $\l$ in $\tilde V$
have a scale $n_\l\ge n_V$.

For each resonance $V$ of $\th^1$ we shall call $M_V$ the number of
vertices in $\tilde V$.  To the just defined set of diagrams we add the
diagrams obtained by reversing simoultaneously the signs of the vertex
modes $\nn_w$, for
$w\in \tilde V$\footnote{${}^3$}{\nota
This can be done without breaking
the relationship which has to exist between the lines, as it can
be easily checked by observing that $ \sum_{w \in \tilde V}
\nn_w = \vec0 $, since, for any resonance $V$,
$ \sum_{v \in V} \nn_v = \vec0 $. }: the change of sign is performed
independently for the various resonant clusters.  This defines a family
of $\prod 2M_V$ diagrams that we call $\FF(\th_1)$. The number
$\prod 2M_V$ will be bounded by $\exp\sum2M_V\le e^{2k}$.

Let $\l$ be a line, in a cluster $T$, contained inside the resonances
$V=V_1\subset V_2\subset\ldots$ of scales $n=n_1>n_2>\ldots$; then the
shifting of the lines $\l_{V_i}$ can cause a change in the size
of the propagator of $\l$ by at most $\g_{n_1}+\g_{n_2}+\ldots
< 2^{n_1} + 2^{n_2} + \ldots < 2^{n+1}$.

Since the number of lines inside $V$ is smaller than $\lis N_n
\equiv E 2^{-n \t^{-1}}$, ($E=2^{-3 \t^{-1}} N^{-1}$), the
quantity $\oo\cdot\nn_\l$ of $\l$ has the form
$\oo\cdot\nn^0_\l+\s_\l\oo\cdot\nn_{\l_V}$ if $\nn^0_\l$ is the momentum
of the line $\l$ ``inside the resonance $V$'',
\ie it is the sum of all the vertex modes of the vertices
preceding $\l$ in the sense of the
line arrows, but contained in $V$; and $\s_\l=0,\pm1$.

Therefore not only $|\oo\cdot\nn^0_\l| > 2^{n+3}$ (because $\nn^0_\l$
is a sum  of $\le \lis N_n$ vertex modes, so that $|\nn^0_\l|\le N\lis
N_n$) but $\oo\cdot\nn^0_\l$ is ``in the middle''
of the interval of scales containing it and, by the
proposition in section 3 (in [G] this was
a consequence of the strong diophantine
property), does not get out of it if we add a quantity
bounded by $2^{n+1}$ (like $\s_\l\oo\cdot\nn_{\l_V}$).
{\it Hence no line changes scale as $\th$ varies in $\FF(\th^1)$}.

Let $\th^2$ be a diagram not in $\FF(\th^1)$ and construct
$\FF(\th^2)$, \etc.  We define a collection
$\{\FF(\th^i)\}_{i=1,2,\ldots}$ of pairwise disjoint families of
diagrams.  We shall sum all the contributions to $\V h^{(k)}$ coming
from the individual members of each family.  This is similar to
the {\it Eliasson's resummation}.

We call $\e_V$ the quantity $\oo\cdot\nn_{\l_V}$ associated with the
resonance $V$ with scale $n$. If $\l$ is a line in $\tilde V$,
(see paragraphs following \equ(4.1)),
we can imagine to write the quantity $\oo\cdot\nn_\l$ as
$\oo\cdot\nn^0_\l+\s_\l\e_V$, with $\s_\l=0,\pm1$.

We want to show that the product of the propagators is
holomorphic in $\e_V$ for $|\e_V|< \g_{n_V-3}$.
Let us reason in the following way.
If $\l$ is a line on scale $n_V$, $\g_{n_V} >
|\oo\cdot\nn_\l| > \g_{n_V-1} $; remarking that it is
$|\oo\cdot\nn^0_\l| > 2^{n+3}$, we obtain immediately
$ | \oo\cdot\nn_\l| >
2^{n+3} - 2^n >  2^{n+2}$, so that $n_V \ge n+3$.
On the other hand, if $n_V > n+3$, \ie $n_V = n+m$, for
some $m>3$,
we note that $ |\oo\cdot\nn^0_\l| > \g_{n_V-1}-\g_n $, because
the resonance scales and the scales of the resonant
clusters (and of all the lines) do not change,
so that it follows that, for $| \e_V | < \g_{n_V-3}$,
$ \, |\oo \cdot \nn_\l^0 + \s_\l \e_V |
\ge \g_{n_V-1}-\g_n-\g_{n_V-3} $
$ \ge (2^{n_V-2} - 2^{n_V-m}) -\g_{n_V-3} $
$ \ge (2^{n_V-3} + 2^{n_V-4} + \ldots + 2^{n_V-m+1}) - 2^{n_V-3} $
$ \ge 2^{n_V-4}$; otherwise,
if $n_V=n+3$, we note that
$|\oo \cdot \nn_\l^0 | > 2^{n+3}$, so that
$ |\oo \cdot \nn_\l^0 + \s_\l \e_V |
> 2^{n+3} - \g_{n_V-3} $ $\ge 2^{n_V-1}$,
for $| \e_V | < \g_{n_V-3}$. Therefore we can conclude that,
while $\e_V$ varies in a complex disk
of radius $\g_{n_V-3}$ and center in $0$,
the quantity $|\oo\cdot\nn^0_\l+\s_\l\e_V|$
does not become smaller than $2^{n_V-4}$.
Note the main point here: the
quantity $\g_{n_V-3}$ will usually be $\gg \g_{n_{\l_V}}$ which is the
value $\e_V$ actually can reach in every diagram in $\FF(\th^1)$; this
can be exploited in applying the maximum principle, as done below.

It follows that, calling $n_\l$ the scale of the line $\l$ in $\th^1$,
each of the $\prod 2 M_V\le e^{2k}$ products of propagators
of the members of the family $\FF(\th^1)$ can be bounded above by
$ \prod_\l\,2^{-2(n_\l-4)}=2^{8k}\prod_\l\,2^{-2n_\l}$, if regarded as a
function of the quantities $\e_V=\oo\cdot\nn_{\l_V}$, for $|\e_V|\le
\,\g_{n_V-3}$, associated with the resonant clusters $V$.  This even
holds if the $\e_V$ are regarded as independent complex parameters.

By construction it is clear that the sum of the $\prod 2M_V\le e^{2k}$
terms, giving the contribution to $\V h^{(k)}$ from the trees in
$\FF(\th^1)$, vanishes to second order in the $\e_V$ parameters (by the
approximate cancellation discussed above).  Hence by the maximum
principle, and recalling that each of the scalar products in \equ(8) can
be bounded by $N^2$, we can bound the contribution from the family
$\FF(\th^1)$ by:
$$ \left[\fra1{k!} \Big(\fra{f_0 2^{2\t}
C_0^2 N^2}{J_0}\Big)^k 2^{8k} e^{2k}
\prod_{n\le0}2^{-2nN_n}\right]\left[\prod_{n\le0}\prod_{T,\,n_T=n}
\prod_{i=1}^{m_T}\,2^{2(n-n_{i}+4)}\right] \Eq(4.2) $$
where:
\acapo
1) $N_n$ is the number of propagators of scale $n$ in $\th^1$ ($n=1$
does not appear as $|\oo\cdot\nn| \ge \g_0 \ge 2^2 $,
in such cases, and $ 2^8 \ge 2^4 $);
\acapo
2) the first square bracket is the bound on the product of
individual elements in the family $\FF(\th^1)$ times the bound $e^{2k}$
on their number;
\acapo
3) the second term is the part coming from the maximum principle
(in the form of Schwarz's lemma), applied
to bound the resummations, and is explained as follows:
\acapo
i) the dependence on the variables $\e_{V_i}\=\e_i$ relative to
resonances $V_i\subset T$ with scale $n_{\l_{V_i}}=n$ is holomorphic for
$|\e_i|< \,\g_{ n_i-3}$ if $n_i\=n_{V_i}$, provided $n_i \ge
n+3$ (see above);
\acapo
ii) the resummation says that the dependence on the $\e_i$'s has a
second order zero in each.  Hence the maximum principle tells us that
we can improve the bound given by the first factor in \equ(4.2) by the
product of factors $ (|\e_i| \, \g_{n_i-3}^{-1})^2
\le 2^{2(n - n_i + 4)} $,
if $n_i \ge n+3$ (of course the gain factor can be
important only when $\ll1$).

Hence substituting \equ(4.1) into
\equ(4.2) we see that the $m_T$ is taken
away by the first factor in $\,2^{2n}2^{-2n_{i}}$, while the
remaining $\,2^{-2n_i}$ are compensated by the $-1$ before
the $+m_T$ in \equ(4.1), taken from the factors with $T=V_i$
(note that there are always enough $-1$'s).

Hence the product \equ(4.2) is bounded by:
$$ \fra1{k!}\,(2^{2\t} C_0^2 J_0^{-1}f_0 N^2)^k e^{2k}2^{8k}2^{8k}
\prod_n\,2^{-8 n k E^{-1}\,2^{\e n}}\le \fra1{k!} B_0^k
\Eq(4.3) $$
with $B_0 = 2^{18} e^2 ( 2^{2\t} C_0^2 f_0 J_0^{-1} ) N^2
\exp [ 2^{3 \t^{-1}} ( 8 N \log 2 ) \sum_{n=1}^{\io}
n 2^{-n \t^{-1} } ] $.
To sum over the trees we note that, fixed $\th$ the collection of
clusters is fixed.  Therefore we only have to multiply \equ(4.3) by the
number of diagram shapes for $\th$, ($\le 2^{2k}k!$), by the number of
ways of attaching mode labels, ($\le (3N)^{lk}$), so that we can bound
$|h^{(k)}_{\nn j}|$ by \equ(1.5).
\*

\penalty-200\vskip1.truecm

\\{\bf Appendix : Resonant Siegel-Brjuno bound.}
\penalty10000\vskip0.5truecm

Calling $N^*_n$ the number of non resonant lines carrying a scale label
$\le n$. We shall prove first that $N^*_n\le 2k (E 2^{-\e n})^{-1}-1$ if
$N_n>0$. We fix $n$ and denote $N^*_n$ as $N^*(\th)$.

If $\th$ has the root line with scale $>n$ then calling
$\th_1,\th_2,\ldots,\th_m$ the subdiagrams of $\th$ emerging from the
first vertex of $\th$ and with $k_j>E\,2^{-\e n}$ lines, it is
$N^*(\th)=N^*(\th_1)+\ldots+N^*(\th_m)$ and the statement is inductively
implied from its validity for $k'<k$ provided it is true that
$N^*(\th)=0$ if $k<E2^{-\e n}$, which is is certainly the case if $E$ is
chosen as in \equ(4.11)\footnote{${}^4$}{\nota Note that if $k\le
E\,2^{-n\e}$ it is, for all momenta $\nn$ of the lines, $|\nn|\le N E
\,2^{-n\e}$, \ie $|\oo\cdot\nn|\ge(NE\,2^{-n\e})^{-\t}=2^3\,2^{n}$ so
that there are {\it no} clusters $T$ with $n_T=n$ and $N^*=0$.}.

In the other case it is $N^*_n\le 1+\sum_{i=1}^mN^*(\th_i)$, and if
$m=0$ the statement is trivial, or if $m\ge2$ the statement is again
inductively implied by its validity for $k'<k$.

If $m=1$ we once more have a trivial case unless the order $k_1$ of
$\th_1$ is $k_1>k-\fra12 E\,2^{-n\e}$.  Finally, and this is the real
problem as the analysis of a few examples shows, we claim that in the
latter case the root line of $\th_1$ is either a resonant line or it has
scale $>n$.

Accepting the last statement it will be: $N^*(\th)=1+N^*(\th_1)=
1+N^*(\th'_1)+\ldots+N^*(\th'_{m'})$, with $\th'_j$ being the $m'$
subdiagrams emerging from the first node of $\th'_1$ with orders
$k'_j>E\,2^{-\e n}$: this is so because the root line of $\th_1$ will
not contribute its unit to $N^*(\th_1)$.  Going once more through the
analysis the only non trivial case is if $m'=1$ and in that case
$N^*(\th'_1)=N^*(\th^{\prime \prime}_1)
+ \ldots + N^*( \th^{\prime \prime}_{m^{\prime \prime} })$,
\etc., until we reach a
trivial case or a diagram of order $\le k-\fra12 E\,2^{-n\e}$.

It remains to check that if $k_1>k-\fra12E\,2^{-n\e}$ then the root line of
$\th_1$ has scale $>n$, unless it is entering a resonance.

Suppose that the root line of $\th_1$ has scale $\le n$ and is not
entering a resonance. Note
that $|\oo\cdot\nn(v_0)|\le\,\g_n,|\oo\cdot\nn(v_1)|\le
\,\g_n$, if $v_0,v_1$ are the first vertices of $\th$ and $\th_1$
respectively.  Hence $\d\=|(\oo\cdot(\nn(v_0)-\nn(v_1))|\le2\,2^n$ and
the diophantine assumption implies that $|\nn(v_0)-\nn(v_1)|>
(2\,2^n)^{-\t^{-1}}$, or $\nn(v_0)=\nn(v_1)$.  The latter case being
discarded as $k-k_1<\fra12E\,2^{-n\e}$ (and we are not considering the
resonances: note also that in such case the lines in $\th/\th_1$
different from the root of $\th$ must be inside a cluster), it follows
that $k-k_1<\fra12E\,2^{-n\e}$ is inconsistent: it would in fact imply
that $\nn(v_0)-\nn(v_1)$ is a sum of $k-k_1$ vertex modes and therefore
$|\nn(v_0)-\nn(v_1)|< \fra12NE\,2^{-n\e}$ hence $\d>2^3\,2^n$ which is
contradictory with the above opposite inequality.

A similar, far easier, induction can be used to prove that if $N^*_n>0$
then the number $p$ of clusters of scale $n$ verifies the bound $p\le 2k
\,(E2^{-\e n})^{-1}-1$.  In fact this is true for $k\le E2^{-\e n}$,
(see footnote 4). Let, therefore, $p(\th)$ be the number of clusters of
scale $n$: if the first tree node $v_0$ is not in a cluster of scale $n$
it is $p(\th)=p(\th_1)+\ldots+p(\th_m)$, with the above notation, and
the statement follows by induction.  If $v_0$ is in a cluster of scale
$n$ we call $\th_1$, $\ldots$, $\th_m$ the subdiagrams emerging from the
cluster containing $v_0$ and with orders $k_j> E2^{-\e n}$. It will be
$p(\th)=1+p(\th_1)+\ldots+p(\th_m)$. Again we can assume that $m=1$, the
other cases being trivial. But in such case there will be only one
branch entering the cluster $V$ of scale $n$ containing $v_0$ and it
will have a momentum of scale $\le n-1$. Therefore the cluster $V$ must
contain at least $E2^{-\e n}$ nodes. This means that $k_1\le k-E2^{-\e
n}$: thus \equ(4.1) is proved.

\penalty-200
\*

\vskip0.5truecm

\penalty-200

{\bf References}

\vskip0.5truecm

\penalty10000

\item{[A] } Arnold, V.: {\it Proof of a A.N. Kolmogorov theorem on
conservation of conditionally periodic motions under small perturbations
of the hamiltonian function}, Uspeki Matematicheskii Nauk, 18, 13-- 40,
1963.

\item{[B] } Brjuno, A.: {\it The analytic form of differential
equations}, I: Transactions of the Moscow Mathematical Society, 25,
131-- 288, 1971; and II: 26, 199--239, 1972.

\item{[CF] } Chierchia, L., Falcolini, C.: {\it A direct proof
of the Theorem by Kolmogorov in Hamiltonian systems},
preprint, archived in  {\tt mp\_arc@math.utexas.edu}, \#93-171.

\item{[E] } Eliasson, L. H.: {\it Hamiltonian systems with linear normal
form near an invariant torus}, ed. G. Turchetti, Bologna Conference,
30/5 to 3/6 1988, World Scientific, 1989. And: {\it Generalization of an
estimate of small divisors by Siegel}, ed. E. Zehnder, P. Rabinowitz,
book in honor of J. Moser, Academic press, 1990. But mainly:
{\it Absolutely convergent series expansions
for quasi--periodic motions}, report 2--88, Dept. of Math., University of
Stockholm, 1988.

\item{[G] } Gallavotti, G.: {\it Twistless KAM tori.},
archived in  {\tt mp\_arc@math.utexas.edu}, \#93-172. See also:
Gallavotti, G.: {\it Invariant tori: a field theoretic point
of view on Eliasson's work}, talk at the meeting in
honor of the 60-th birthday of G. Dell'Antonio
(Capri, may 1993), Marseille, CNRS-CPT, preprint

\item{[G2] } Gallavotti, G.: {\it Twistless KAM tori, quasi flat
homoclinic intersections, and other cancellations in the perturbation
series of certain completely integrable hamiltonian systems.  A
review.}, deposited in the archive {\tt mp\_arc@math.utexas.edu},
\#93-164.

\item{[G3] } Gallavotti, G.: {\it The elements of Mechanics}, Springer, 1983.

\item{[G4] } Gallavotti, G.: {\it Renormalization theory and ultraviolet
stability for scalar fields via renormalization group methods}, Reviews
in Modern Physics, 57, 471- 572, 1985. See also: Gallavotti, G.:
{\it Quasi integrable mechanical systems}, Les Houches, XLIII (1984),
vol. II, p. 539-- 624, Ed. K. Osterwalder, R. Stora, North Holland, 1986.

\item{[K] } Kolmogorov, N.: {\it On the preservation of conditionally
periodic motions}, Doklady Aka\-de\-mia Nauk SSSR, 96, 527-- 530, 1954. See
also: Benettin, G., Galgani, L., Giorgilli, A., Strelcyn, J.M.:
{\it A proof of Kolmogorov theorem on invariant tori using canonical
transormations defined by the Lie method}, Nuovo Cimento, 79B, 201--
223, 1984.

\item{[M] } Moser, J.: {\it On invariant curves of an area preserving
mapping of the annulus}, Nach\-rich\-ten Akademie Wiss. G\"ottingen, 11,
1-- 20, 1962.

\item{[P] } P\"oschel, J.: {\it Invariant manifolds of complex analytic
mappings}, Les Houches, XLIII (1984), vol. II, p. 949-- 964, Ed. K.
Osterwalder, R. Stora, North Holland, 1986.

\item{[PV] } Percival, I., Vivaldi, F.: {\it Critical dynamics and
diagrams}, Physica, D33, 304-- 313, 1988.

\item{[S] } Siegel, K.: {\it Iterations of analytic functions}, Annals
of Mathematics, {\bf 43}, 607-- 612, 1943.

\penalty10000

\item{[T] } Thirring, W.: {\it Course in Mathematical Physics}, vol. 1,
p. 133, Springer, Wien, 1983.

\ciao